\documentclass[11pt,paper=a4]{scrartcl}

\usepackage{listings}
\usepackage[dvipsnames]{xcolor}
\usepackage{pifont}
\usepackage{amsmath}
\usepackage{bm}
\usepackage[margin=2.5cm,includehead=true,includefoot=true,centering,]{geometry}
\usepackage{footnotebackref}
\usepackage[sort&compress, numbers]{natbib}
\usepackage[semibold, scale=0.9]{sourcecodepro}
\usepackage{dirtree}
\usepackage{multicol}
\usepackage{graphicx}
\usepackage{authblk}
\usepackage{multirow}
\usepackage{markdown}
\usepackage{fontawesome5}

\setkomafont{disposition}{\normalfont\bfseries}
\DeclareOldFontCommand{\bf}{\normalfont\bfseries}{\mathbf}

\setcapindent{0em}

\usepackage{hyperref}
\hypersetup{                   %
    colorlinks = true,         
    linkcolor  = MidnightBlue, 
    urlcolor   = MidnightBlue,     
    citecolor  = MidnightBlue, 
    pdfcreator={LaTeX via pandoc with the Eisvogel template}
}

\makeatletter
\AtBeginDocument{%
  \newcounter{llabel}[lstlisting]%
  \renewcommand*{\thellabel}{%
    \ifnum\value{llabel}<0 %
      \@ctrerr
    \else
      \ifnum\value{llabel}>10 %
        \@ctrerr
      \else
        \protect\ding{\the\numexpr\value{llabel}+201\relax}%
      \fi
    \fi
  }%
}

\newlength{\llabelsep}
\newcommand*{\llabel}[1]{%
  \begingroup
    \refstepcounter{llabel}%
    \label{#1}%
    \llap{%
      \thellabel\kern\llabelsep
      \hphantom{\lst@numberstyle\the\lst@lineno}%
      \kern\lst@numbersep
    }%
  \endgroup
}
\makeatother

\usepackage{caption}
\DeclareCaptionFont{white}{\color{white}}
\DeclareCaptionFont{white}{\color{white}}
\DeclareCaptionFormat{listing}{\colorbox{gray}{\parbox{\dimexpr\linewidth-2\fboxsep\relax}{#1#2#3}}}
\definecolor{listing-background}{HTML}{F4F4F4}
\definecolor{listing-background-terminal}{HTML}{000000}
\definecolor{listing-rule}{HTML}{B3B2B3}
\definecolor{listing-numbers}{HTML}{B3B2B3}
\definecolor{listing-text-color}{HTML}{000000}
\definecolor{listing-text-color-terminal}{HTML}{FFFFFF}
\definecolor{listing-keyword}{HTML}{1284CA}
\definecolor{listing-keyword-2}{HTML}{1800AA} 
\definecolor{listing-keyword-3}{HTML}{9137CB} 
\definecolor{listing-identifier}{HTML}{435489}
\definecolor{listing-string}{HTML}{00999A}
\definecolor{listing-comment}{HTML}{8E8E8E}

\definecolor{draculabg}      {RGB} {40,   42,   54}
\definecolor{draculacl}      {RGB} {68,   71,   90}
\definecolor{draculafg}      {RGB} {248,  248,  242}
\definecolor{draculacomment} {RGB} {98,   114,  164}
\definecolor{draculacyan}    {RGB} {139,  233,  253}
\definecolor{draculagreen}   {RGB} {24,   182,  82}
\definecolor{draculaorange}  {RGB} {250,  155,  75}
\definecolor{draculapink}    {RGB} {255,  121,  198}
\definecolor{draculapurple}  {RGB} {189,  147,  249}
\definecolor{draculared}     {RGB} {255,  85,   85}
\definecolor{draculayellow}  {RGB} {241,  250,  140}

\definecolor{pdocbg}{RGB}{248, 248, 248} 
\definecolor{pdockeyword}{RGB}{0, 0, 204} 
\definecolor{pdocstring}{RGB}{34, 139, 34} 
\definecolor{pdoccomment}{RGB}{128, 128, 128} 
\definecolor{pdocnumber}{RGB}{128, 128, 128} 

\lstdefinelanguage{mypython}{
language=Python,
keywordstyle     = [1]{\color{draculapurple}\bfseries},
keywordstyle     = [2]{\color{draculaorange}},
keywordstyle     = [3]{\color{draculapink}},
keywordstyle     = [6]{\color{draculared}\bfseries\itshape},
identifierstyle  = \color{draculacl},
commentstyle     = \color{draculacomment},
stringstyle      = \color{draculared},
showstringspaces = false,
morekeywords     = [1]{as, assert, nonlocal, with, yield, self, True, False, None}, 
morekeywords=[2]{__init__,__add__,__mul__,__div__,__sub__,__call__,__getitem__,__setitem__,__eq__,__ne__,__nonzero__,__rmul__,__radd__,__repr__,__str__,__get__,__truediv__,__pow__,__name__,__future__,__all__}, 
morekeywords=[2]{object,type,isinstance,copy,deepcopy,zip,enumerate,reversed,list,set,len,dict,tuple,range,xrange,append,execfile,real,imag,reduce,str,repr}, 
morekeywords=[2]{ode,fsolve,sqrt,exp,sin,cos,arctan,arctan2,arccos,pi, array,norm,solve,dot,arange,isscalar,max,sum,flatten,shape,reshape,find,any,all,abs,plot,linspace,legend,quad,polyval,polyfit,hstack,concatenate,vstack,column_stack,empty,zeros,ones,rand,vander,grid,pcolor,eig,eigs,eigvals,svd,qr,tan,det,logspace,roll,min,mean,cumsum,cumprod,diff,vectorize,lstsq,cla,eye,xlabel,ylabel,squeeze}, 
morekeywords=[2]{compile_model, yscale, ylim, show, save_variables, save_spec, load_GEFdata, read_spec, omega_gw, compute_pt, run},
morekeywords=[3]{GaugeSpec, PowSpecT, paiGEF, faiGEF,},
morekeywords=[6]{Exception,NameError,IndexError,SyntaxError,TypeError,ValueError,OverflowError,ZeroDivisionError}, 
}

\lstdefinestyle{pdoc}{
    basicstyle=\ttfamily\footnotesize,
    keywordstyle=\color{pdockeyword}\ttfamily,
    stringstyle=\color{pdocstring}\ttfamily,
    commentstyle=\color{pdoccomment}\ttfamily,
    backgroundcolor=\color{pdocbg},
    frame=single,
    rulecolor=\color{gray!30},
    numbers=left,
    numberstyle=\tiny\color{pdocnumber},
    breaklines=true,
    showstringspaces=false,
    captionpos=b,
    belowcaptionskip=1\baselineskip,
    xleftmargin=0.5cm,
    framexleftmargin=0.5cm,
  }

\definecolor{mygreen}{RGB}{31, 138, 112}

\newcommand{\diver}{\operatorname{div}}
\newcommand{\D}{\mathrm{d}}
\newcommand{\MP}{M_\mathrm{P}}
\newcommand{\geff}{\texttt{GEFF}}
\newcommand{\code}[1]{\texttt{#1}}

\lstdefinelanguage{mybash}{
language=bash,
keywordstyle     = {\color{draculapurple}\bfseries},
keywordstyle     = [1]{\color{draculapink}\bfseries},
keywordstyle     = [2]{\color{draculapurple}},
keywordstyle     = [3]{\color{draculaorange}},
morekeywords     = [1]{pull, run, create, install,notebook, shell},
morekeywords     = [2]{docker, singularity, conda, bash, pip, brew, curl, sudo, apt,get,jupyter},
}
\lstdefinestyle{eisvogel_listing_style}{
  numbers          = left,
  xleftmargin      = 2.25em,
  xrightmargin      = \fboxsep,
  framexleftmargin = 2em,
  backgroundcolor  = \color{listing-background},
  basicstyle       = \color{draculacl}\linespread{1.0}%
                      \ttfamily{},
  numberstyle       = \color{draculacl}\small%
                      \ttfamily{},
  escapeinside     = {/*@}{@*/}, 
  breaklines       = true,
  sensitive        = true,
  frame            = single,
  framesep         = 0.19em,
  rulecolor        = \color{listing-rule},
  frameround       = ffff,
  tabsize          = 4,
  numbers          = none,
  aboveskip        = 1.0em,
  belowskip        = 0.5em,
  abovecaptionskip = 0em,
  belowcaptionskip = 0.75em,
  literate         =
  {á}{{\'a}}1 {é}{{\'e}}1 {í}{{\'i}}1 {ó}{{\'o}}1 {ú}{{\'u}}1
  {Á}{{\'A}}1 {É}{{\'E}}1 {Í}{{\'I}}1 {Ó}{{\'O}}1 {Ú}{{\'U}}1
  {à}{{\`a}}1 {è}{{\`e}}1 {ì}{{\`i}}1 {ò}{{\`o}}1 {ù}{{\`u}}1
  {À}{{\`A}}1 {È}{{\`E}}1 {Ì}{{\`I}}1 {Ò}{{\`O}}1 {Ù}{{\`U}}1
  {ä}{{\"a}}1 {ë}{{\"e}}1 {ï}{{\"i}}1 {ö}{{\"o}}1 {ü}{{\"u}}1
  {Ä}{{\"A}}1 {Ë}{{\"E}}1 {Ï}{{\"I}}1 {Ö}{{\"O}}1 {Ü}{{\"U}}1
  {â}{{\^a}}1 {ê}{{\^e}}1 {î}{{\^i}}1 {ô}{{\^o}}1 {û}{{\^u}}1
  {Â}{{\^A}}1 {Ê}{{\^E}}1 {Î}{{\^I}}1 {Ô}{{\^O}}1 {Û}{{\^U}}1
  {œ}{{\oe}}1 {Œ}{{\OE}}1 {æ}{{\ae}}1 {Æ}{{\AE}}1 {ß}{{\ss}}1
  {ç}{{\c c}}1 {Ç}{{\c C}}1 {ø}{{\o}}1 {å}{{\r a}}1 {Å}{{\r A}}1
  {€}{{\EUR}}1 {£}{{\pounds}}1 {«}{{\guillemotleft}}1
  {»}{{\guillemotright}}1 {ñ}{{\~n}}1 {Ñ}{{\~N}}1 {¿}{{?`}}1
  {…}{{\ldots}}1 {≥}{{>=}}1 {≤}{{<=}}1 {„}{{\glqq}}1 {“}{{\grqq}}1
  {”}{{''}}1
}
\usepackage{tikz}

\begin{document}

\title{\geff \\\href{https://richard-von-eckardstein.github.io/GEFF/}{\color{black}The Gradient Expansion Formalism Factory}\\ 
           {\Large A tool for inflationary gauge-field production}\\{\Large v0.1.0}}

\hyphenation{bootstrap}
\hyphenation{ptarcade}

\date{}

\setlength{\affilsep}{-0.05em}
\author{
\textbf{ Main developer:}\vspace{-1em}\and
{\normalsize Richard von Eckardstein}\\ 
{\footnotesize{\emph{Institute for Theoretical Physics, University of M\"unster,}}}\\
\vspace{-0.5em}{{\footnotesize{\emph{Wilhelm-Klemm-Stra{\ss}e 9,  48149 M\"{u}nster, Germany}}}}\\\smallskip
{{\footnotesize{E-mail: \href{mailto:richard.voneckardstein@uni-muenster.de}{richard.voneckardstein@uni-muenster.de}} }}
}
\maketitle

\begin{tikzpicture}[remember picture,overlay]
\node[anchor=north east,inner sep=30pt] at (current page.north east) {MS-TP-25-37};
\end{tikzpicture}

\begin{abstract}
    The \geff\ --- the \textit{Gradient Expansion Formalism Factory} --- is a new \texttt{Python} package designed to study gauge-field production during inflation.
    The package provides a framework to implement and use the gradient expansion formalism (GEF), a numerical technique devised to study the nonlinear dynamics associated with inflationary gauge-field generation.
    The GEF has already been applied in the context of axion inflation, and with the \geff\ package, one can build on these results.
    The \geff\ gives users access to ready-to-use model files for two scenarios of axion inflation: pure axion inflation, with the inflaton coupled to a pure Abelian gauge sector, 
    and fermionic axion inflation, which assumes that the Standard Model (SM) hypercharge field is coupled to the inflaton, resulting in the production SM fermions via the Schwinger effect.
    The \geff\ provides the user with methods to solve GEF equations, including an integrated error estimator and self-correction algorithm.
    Furthermore, users can implement their own GEF models, e.g., variations of axion inflation or related scenarios.
    The package also comes with tools to study the production of primordial gravitational waves induced by gauge fields.
    This is a starting guide for the \geff\, providing a high-level introduction to the GEF, installation instructions, and the basics for using this package.
    \href{https://github.com/richard-von-eckardstein/GEFF}{\color{black}\faGithub}
\end{abstract}
\thispagestyle{empty}
\clearpage

\section*{Program summary}
\textit{Program Title:} GEFF \smallskip

\noindent \textit{Licensing provisions:} MIT \smallskip

\noindent \textit{Programming language:} Python \smallskip

\noindent \textit{GitHub repository:} \href{https://github.com/richard-von-eckardstein/GEFF}{\color{black}https://github.com/richard-von-eckardstein/GEFF} \smallskip

\noindent \textit{Documentation:} \href{https://richard-von-eckardstein.github.io/GEFF}{\color{black}https://richard-von-eckardstein.github.io/GEFF} \smallskip

\noindent \textit{Nature of Problem:} Inflationary gauge-field production can be an efficient process, and can induce strong backreaction effects onto the dynamics of inflation.
The inflationary dynamics in the presence of gauge-field backreaction are highly nonlinear and require numerical techniques to solve for them.
While lattice simulations are suitable for this task, they are computationally expensive and are limited in their dynamical range.
An alternative is the gradient expansion formalism (GEF)~\cite{Sobol:2020lec,Gorbar_2021}. In this approach, gauge-field evolution is formulated as an infinite tower of coupled ordinary differential equations (ODEs)
that may be approximately closed at some order $n_\mathrm{tr}$. In this way, the GEF provides a numerically fast alternative to handle gauge-field backreaction. Solutions obtained with the GEF can be used on their own, or to establish \textit{GEF benchmark} to inform more costly 
lattice simulations. However, the truncation of the infinite ODE tower can lead to a breakdown of the GEF technique. This requires algorithms that combine the GEF method with a momentum-space approach to ensure a trustable result.\smallskip

\noindent \textit{Solution method:} The \geff\ is designed to handle the infinite ODE towers appearing in the GEF. This includes an algorithm that uses GEF solutions in concert with a momentum-space result to 
detect a breakdown of the ODE truncation scheme, and attempt to self-correct the GEF in case a breakdown has occurred. The package comes with ready-to-use models for the dynamics of axion inflation, and 
supports the user in creating custom models adapted to other scenarios.
\thispagestyle{empty}
\clearpage
\pagenumbering{arabic}

\tableofcontents

\section{Introduction}

Inflationary gauge-field production gives rise to a plethora of fascinating phenomenological consequences.
    A primary motivation for studying primordial gauge fields is inflationary magnetogenesis~\cite{Turner_1987,Ratra_1991,garretson_1992}, which could help to explain the evidence for large-scale magnetic fields in cosmic voids
    (see also Refs.~\cite{Durrer:2013pga,Subramanian:2015lua,Vachaspati:2020blt} for reviews).
    Furthermore, strong gauge fields present during inflation will imprint themselves onto the scalar and tensor power spectrum~\cite{Yokoyama:2008xw, Anber_2010,Dimastrogiovanni:2010sm,Barnaby_2011_A,Barnaby_2011_B,Sorbo_2011}.
    This way, gauge fields can contribute to an inflationary stochastic gravitational-wave background~\cite{Barnaby_2011,cook_2012}, 
    or lead to the formation of primordial black holes~\cite{linde_2013}.
    The transfer of energy to gauge fields can also be an efficient process for preheating~\cite{adshead_2015},
    while the decay of helical inflationary gauge fields is relevant for baryogenesis~\cite{anber_2015}.

    A strong inflationary electromagnetic field also efficiently sources charged particles coupled to it via Schwinger pair creation~\cite{Frob_2014,Kobayashi_2014}.
    The conductive medium composed of these charge carriers can backreact on the gauge fields by dampening them~\cite{Sobol_2018,Sobol_2019}.
    At the same time, this source of charged fields could also provide a mechanism to explain the abundance of dark matter~\cite{bastero-gil_2024_A}.

To source gauge fields during inflation, one needs to break conformal invariance~\cite{Parker_1968}. 
For example, this can be achieved by coupling them to a scalar field $\varphi$, either via a kinetic coupling, $I_K(\varphi) F_{\mu\nu} F^{\mu\nu}$~\cite{Ratra_1991}, 
or an axial coupling, $I_A(\varphi) F_{\mu\nu} \tilde{F}^{\mu\nu}$~\cite{garretson_1992}, with $F_{\mu\nu}$ the field-strength tensor and $\tilde{F}_{\mu\nu}$ its dual. 
Of particular interest is the case when $\varphi$ is the inflaton. For an axial coupling, such a model is referred to as axion inflation~\cite{garretson_1992,anber_2006, Anber_2010} (see also Ref.~\cite{Pajer_2013} for a review). 
In dilaton models, the inflaton instead couples to the kinetic term~\cite{Ratra_1991,Gasperini:1995dh}.
However, the scalar field could also be a spectator during inflation, e.g., in models of axion spectator inflation~\cite{Namba_2015}.
Furthermore, while much of the available literature focuses on Abelian gauge fields, non-Abelian fields during inflation are also considered; see, e.g., Refs~\cite{Maleknejad:2011jw,Maleknejad:2011sq,Adshead_2012}.

If the production of gauge bosons is efficient, the resulting electromagnetic fields can backreact on the inflationary dynamics giving rise to highly nonlinear dynamics (for example, Refs.~\cite{
    Cheng_2016,Domcke_2020_Resonant,Sobol:2020lec,Gorbar_2021,caravano_2022,eckardstein_2023,
gorbar_2023,Gorbar:2023zla,Durrer:2023rhc,caravano_2023,Figueroa_2023,Figueroa_2024,eckardstein_2025,sharma_2025,vonEckardstein_PAI_2025,vonEckardstein:2025elq}).
To model these dynamics, several techniques have been proposed in the literature, one of which is the gradient expansion formalism (GEF)~\cite{Sobol:2020lec,Gorbar_2021}.
 In this method, one directly evolves bilinear expectation values of electric and magnetic fields, $\langle \bm{E}^2 \rangle$,  $\langle \bm{B}^2 \rangle$,  $\langle \bm{E} \cdot \bm{B} \rangle$, etc.,
 as backreaction onto the inflationary spacetime typically proceeds via these terms. This technique has already been applied to axion inflation~\cite{Gorbar_2021,eckardstein_2023,gorbar_2023,Gorbar:2023zla,domcke_2024,eckardstein_2025,vonEckardstein_PAI_2025,vonEckardstein:2025elq} and
  dilatonic models~\cite{Sobol:2020lec,Durrer:2023rhc,Lysenko:2025pse,Lysenko:2025sdo}.
 Furthermore, the GEF can capture the effect of Schwinger pair creation by effectively modelling the fermions as a conductive medium which dampens gauge-field generation~\cite{Gorbar_2021,gorbar_2023,eckardstein_2025}.

The \geff --- the \textit{Gradient Expansion Formalism Factory} ---  is a novel \code{Python} package that allows the user 
to harness the power of the GEF to study the diverse implications of inflationary gauge-field generation.
The package comes with a built-in version of two axion inflation models.
The first model is pure axion inflation (PAI), which consists of an Abelian gauge field coupled to the inflaton via a linear axial coupling,  $\varphi  F_{\mu\nu} \tilde{F}^{\mu\nu}$.\footnote{
    It should be noted that recent lattice simulations~\cite{caravano_2023,Figueroa_2023,Figueroa_2024,sharma_2025} of pure Abelian axion inflation show that strong gauge-field backreaction is accompanied by the production of large axion inhomogeneities, which alter the dynamics of inflation.
    In its current form, the GEF cannot reliably capture these effects~\cite{domcke_2024}.
    However, the GEF is a fast and efficient technique and can be used to establish a \textit{GEF benchmark} to inform costly lattice simulations~\cite{vonEckardstein_PAI_2025,vonEckardstein:2025elq}.
 }
This model does not account for Schwinger pair production.
The second model, fermionic axion inflation (FAI), extends PAI to properly capture the case when the Abelian gauge field is identified with the hypercharge field of the Standard Model (SM) by
accounting for Schwinger pair production of SM fermions. The FAI model implemented in the \geff\ captures the effect of SM fermions by treating them as an effective conductive medium which dampens gauge-field production.
For details on these models, see Refs.~\cite{Gorbar_2021,eckardstein_2025}.
In addition to these built-in models, the \geff\ also provides the user with the option to implement their own GEF model by following a simple set of rules for writing a model file.
Lastly, the \geff\ comes with tools tailored toward the \geff\ output. Currently, this involves a routine to compute the tensor power spectrum induced by gauge fields and the corresponding
spectrum of gravitational waves. We plan on extending these tools in the future. For more details on the \geff\ see the \href{https://richard-von-eckardstein.github.io/GEFF}{documentation}.

This document is a lightweight manual for the \geff\ in which we introduce its basic functionalities. In Section~\ref{sec: primer}, we give a high-level summary of the GEF technique. 
Instructions for installing the \geff\ package are given in Section~\ref{sec: installation}. Section~\ref{sec: tour} provides a working example on how to use this package, taking the built-in version of PAI for illustration.
Lastly, in Section~\ref{sec: pictures}, we showcase some of the results which may be obtained using the \geff.

\section{A primer on the GEF}
\label{sec: primer}

Here, we introduce the GEF in the broadest strokes, without getting into the details.
 For a more complete picture, see, e.g., Refs.~\cite{Gorbar_2021,eckardstein_2023,eckardstein_2025,domcke_2024}.

As a start, consider Maxwell's equations in an expanding spacetime,\footnote{
    We use physical time, $t$, with its derivative denoted as $\dot{f}(t)$. The Hubble rate is $H$ and the scale factor is $a$.
}

\begin{subequations}
    \begin{equation}
        \bm{E} = 0\, , \qquad \operatorname{div} \bm{B} = 0 \, ,
        \label{eq: constraints}
    \end{equation}
    \begin{equation}
        \dot{\bm{E}} + 2 H \bm{E} - \frac{1}{a}\operatorname{rot} \bm{B} + \bm{J} = 0 \, ,
        \label{eq: Ampere}
    \end{equation}
    \begin{equation}
        \dot{\bm{B}}  + 2 H \bm{B} + \frac{1}{a}\operatorname{rot} \bm{E} = 0 \,
        \label{eq: Faraday}
    \end{equation}
\end{subequations}
where $\bm{E}$ and $\bm{B}$ are electric and magnetic field operators, and $\bm{J}$ is linear in $\bm{E}$ and $\bm{B}$.

A typical approach to solving these equations is to express $\bm{E}$ and $\bm{B}$ in terms of mode functions $A_\lambda(t,k)$,\footnote{
    We use radiation gauge, $\diver\vec{A} = 0$ and  $A_0=0$. The ladder operators $ \hat{a}_\lambda(\bm{k})$,  $\hat{a}^{{\dagger}}_\lambda(\bm{k})$ obey the typical commutation relations, and 
    $\bm{\epsilon}_\lambda(\bm{k})$ are polarization vectors in a helicity basis, with $\lambda = \pm1$ labeling the two polarizations.
}

\begin{subequations}
    \begin{equation}
        \bm{E}(t, \bm{x}) = - \frac{1}{a} \int \frac{\D^3 \bm{k}}{(2 \pi)^{3/2}}\sum_{\lambda={\pm1}} 
                \left( \bm{\epsilon}_\lambda(\bm{k}) \dot{A}_\lambda(t, k) \hat{a}^{\vphantom{\dagger}}_\lambda(\bm{k})e^{i \bm{k} \cdot \bm{x}} 
                + \mathrm{h.c.} \right) \, ,
    \end{equation}
    \begin{equation}
        \bm{B}(t, \bm{x}) = \frac{1}{a^2} \int \frac{\D^3 \bm{k}}{(2 \pi)^{3/2}}\sum_{\lambda={\pm1}} \lambda k
                \left(  \bm{\epsilon}_\lambda(\bm{k}) A_\lambda(t, k) \hat{a}^{\vphantom{\dagger}}_\lambda(\bm{k})e^{i \bm{k} \cdot \bm{x}} 
                + \mathrm{h.c.} \right) \, .
    \end{equation}
    \label{eq: E and B}%
\end{subequations}%
In principle, the mode functions $A_\lambda(t,k)$ obey a simple evolution equation,\footnote{We write $J_\lambda(t,k)$ for the Fourier modes of $J$.}
\begin{equation}
    \ddot{A}_\lambda(t,k) + H\dot{A}_\lambda(t,k) + \left(\frac{k}{a}\right)^2 A_\lambda(t,k) = J_\lambda(t,k) \, .
    \label{eq: mode-eq}
\end{equation}
However, since the inflationary background evolution of an FLRW spacetime should depend on homogeneous and isotropic quantities,
any dependence of the background evolution on $\bm{E}$ and $\bm{B}$ will typically be given by terms like, $\langle\bm{E}^2\rangle$, $\langle\bm{B}^2\rangle$, $\langle\bm{E} \cdot \bm{B}\rangle$, etc., where $\langle \cdot \rangle$
is the quantum expectation value with respect to the Bunch--Davies vacuum.
For example, via the Friedmann equation, the Hubble rate will depend on $\langle\bm{E}^2\rangle$ and $\langle\bm{B}^2\rangle$,
\begin{equation}
    H^2 \supset \frac{\langle\bm{E}^2 + \bm{B}^2\rangle}{6 \MP^2} = \frac{1}{{12 \MP^2 \pi^2}}\sum_\lambda \int_0^\mathrm{k_\mathrm{UV}(t)}\!\D k\, \left(\frac{k}{a} \right)^2 \left[ |\dot{A}_\lambda(t,k)|^2 + \left(\frac{k}{a}\right)^2|A_\lambda(t,k)|^2 \right] \, .
\end{equation}
The right-hand side of this expression is challenging to evaluate in a momentum-space approach: 
to determine the Hubble rate, one needs to evaluate $\langle\bm{E}^2\rangle$ and $\langle\bm{B}^2\rangle$, and they depend on an integral over $A_\lambda(t,k)$, whose evaluation itself depends on $H$. 
Even worse, $\langle\bm{E}^2\rangle$ and $\langle\bm{B}^2\rangle$ need to be regulated, and, in a time-dependent background, a regulating momentum cut-off $k_\mathrm{UV}$ will itself depend on time.
Thus, a numerical evaluation is costly; at every time step, an integral has to be evaluated, and the necessary range of modes $k \leq k_\mathrm{UV}(t_\mathrm{max})$, is a priori unknown.

This is where the GEF comes into play. In the GEF, one does not take the detour to momentum space but instead directly evolves $\langle\bm{E}^2\rangle$, $\langle\bm{B}^2\rangle$, $\langle\bm{E} \cdot \bm{B}\rangle$, etc.
To realize this, one introduces the quantities,
\begin{subequations}
    \begin{align}
        \mathcal{F}_\mathcal{E}^{(n)} &= \frac{a^{n+4}}{k_{\mathrm{UV}}^{n+4}}\langle \bm{E} \cdot \operatorname{rot}^n \bm{E}\rangle 
                                = \int\limits_{0}^{k_{\mathrm{UV}}(t)}\frac{\D k}{k} \frac{a^2 k^{n+3}}{2 \pi^2 k_{\mathrm{UV}}^{n+4}}  \sum_{\lambda}\lambda^n |\dot{A}_\lambda(t,k)|^2\, , \\
        \mathcal{F}_\mathcal{G}^{(n)} &= -\frac{a^4}{k_{\mathrm{UV}}^{n+4}}\langle \bm{E} \cdot \operatorname{rot}^n \bm{B}\rangle 
                                = \int\limits_{0}^{k_{\mathrm{UV}}(t)} \frac{\D k}{k} \frac{a k^{n+4}}{2 \pi^2 k_{\mathrm{UV}}^{n+4}}\sum_{\lambda}\lambda^{n+1} \operatorname{Re}[\dot{A}_\lambda(t,k)A_\lambda^*(t,k)] \, , \\
        \mathcal{F}_\mathcal{B}^{(n)} &= \frac{a^4}{k_{\mathrm{UV}}^{n+4}}\langle \bm{B} \cdot \operatorname{rot}^n \bm{B}\rangle 
                                = \int\limits_{0}^{k_{\mathrm{UV}}(t)}\frac{\D k}{k} \frac{k^{n+5}}{2 \pi^{2}k_{\mathrm{UV}}^{n+4}} \sum_{\lambda}\lambda^n |A_\lambda(t,k)|^2 \, ,
    \end{align}%
    \label{eq: GEF quantities}%
\end{subequations}%
where, for completeness, we have also given the expression for $\mathcal{F}_\mathcal{X}^{(n)}$ in terms of the mode functions $A_\lambda(t,k)$.
The quantities $\mathcal{F}_\mathcal{X}^{(n)}$ then obey the following coupled differential equations.\footnote{
    We keep $\bm{J}$ arbitrary, to close the ODEs, it should be a linear combination of $\bm{E}$ and $\bm{B}$.
}
\begin{subequations}
    \begin{equation}
         \frac{\D}{\D t} \mathcal{F}_\mathcal{E}^{(n)} + (4+n)\frac{\D \ln k_\mathrm{UV}}{\D t} \mathcal{F}_\mathcal{E}^{(n)}  + 2\frac{k_\mathrm{UV}}{a}\mathcal{F}_\mathcal{G}^{(n+1)} + 2 \frac{a^4}{k_\mathrm{UV}^{n+4}} \langle \bm{J} \cdot \operatorname{rot}^n \bm{E} \rangle =  S_{\mathcal{E}}^{(n)}\, , 
    \end{equation}
    \begin{equation}
         \frac{\D}{\D t} \mathcal{F}_\mathcal{G}^{(n)} + (4+n)\frac{\D \ln k_\mathrm{UV}}{\D t} \mathcal{F}_\mathcal{G}^{(n)} - \frac{k_\mathrm{UV}}{a}\left(\mathcal{F}_\mathcal{E}^{(n+1)} - \mathcal{F}_\mathcal{B}^{(n+1)}\right) - \frac{a^4}{k_\mathrm{UV}^{n+4}} \langle \bm{J} \cdot \operatorname{rot}^n \bm{B} \rangle= S_{\mathcal{G}}^{(n)}\, , 
    \end{equation}
    \begin{equation}
         \frac{\D}{\D t} \mathcal{F}_\mathcal{B}^{(n)} + (4+n)\frac{\D \ln k_\mathrm{UV}}{\D t} \mathcal{F}_\mathcal{B}^{(n)} - 2\frac{k_\mathrm{UV}}{a}\mathcal{F}_\mathcal{G}^{(n+1)}  =  S_{\mathcal{B}}^{(n)}\, .
    \end{equation}%
    \label{eq: GEF ODEs}%
\end{subequations}%
The terms $S_\mathcal{X}^{(n)}$ are boundary terms arising from the time-dependent regulator $k_\mathrm{UV}$; see, e.g., Ref.~\cite{Gorbar_2021}.
Although these are infinitely many ODE's, one can typically determine an analytical closing condition, such that they may be approximately closed at some order $n_\mathrm{tr}$.

This is at the heart of the GEF: by solving Eqs.~\eqref{eq: GEF ODEs} alongside the equations determining the dynamical background evolution, one forgoes the need for an expensive momentum space computation.
After the evolution of the time-dependent background has been obtained in this way, one can still determine the gauge-field spectra in terms of $A_\lambda(t,k)$ by solving Eq.~\eqref{eq: mode-eq}.
However, this is now straight-forward, as one already knows the time-evolution of $H$, $k_\mathrm{UV}$, etc. 

The \geff\ package is designed to handle an infinite tower of differential equations like Eq.~\eqref{eq: GEF ODEs},
including an algorithm to verify that the analytical approximation for truncating the ODEs at $n_\mathrm{tr}$ does not impact the solution.
From the output, the \geff\ also computes the mode-function evolution, which can then be studied to determine other quantities of interest, such as the gravitational-wave spectrum sourced by the gauge fields.

\section{Installation}
\label{sec: installation}

The \geff-package is available as a PyPI package and can be installed using \href{https://pypi.org/project/pip/}{\code{pip}}:\smallskip
\begin{lstlisting}[language=mybash,numbers=left,mathescape]
pip install cosmo-geff    
\end{lstlisting}

\noindent Alternatively, one can install the \geff\ package in a  \href{https://docs.conda.io/projects/conda/en/latest/index.html}{\code{conda}} environment
by using the \code{geff.yml} file found in \href{https://github.com/richard-von-eckardstein/GEFF}{the GitHub repository} for this package:\smallskip

\begin{lstlisting}[language=mybash,numbers=left,mathescape]
conda env create -f geff.yml 
\end{lstlisting}

\noindent Note that the \geff\ requires Python version 3.10 or higher.

\section{A tour of the factory}
\label{sec: tour}

We will now highlight some of the most important features of the \geff. This is only intended as a complement to the \href{https://richard-von-eckardstein.github.io/GEFF}{documentation}, and should not be taken as a thorough tutorial.

\subsection{The basics}
\label{subsec: basics}

To illustrate the typical workflow using the \geff, we focus on one of its pre-defined models: pure (Abelian) axion inflation (PAI).
In this model, the inflaton $\varphi$ is directly coupled to an Abelian gauge field and acts as an external current, $\bm{J} = (\beta / \MP) \dot{\varphi} \bm{B}$. In the GEF language, the system dynamics are given by
\begin{subequations}
    \begin{equation}
        \ddot{\varphi} + 3 H \dot{\varphi} + V_{,\phi}(\varphi) = - \frac{\beta}{\MP} \left(\frac{a}{k_\mathrm{h}}\right)^4 \mathcal{F}_\mathcal{G}^{(0)}\, ,
        \label{eq: phiEoM}
    \end{equation}
    \begin{equation}
        \frac{\D}{\D t} \mathcal{F}_\mathcal{E}^{(n)} + (4+n)\frac{\D \ln k_\mathrm{h}}{\D t} \mathcal{F}_\mathcal{E}^{(n)}  + 2\frac{k_\mathrm{h}}{a}\mathcal{F}_\mathcal{G}^{(n+1)} - 2 \frac{\beta}{\MP} \dot{\varphi} \mathcal{F}_\mathcal{G}^{(n)}=  S_{\mathcal{E}}^{(n)}\, , 
    \end{equation}
    \begin{equation}
         \frac{\D}{\D t} \mathcal{F}_\mathcal{G}^{(n)} + (4+n)\frac{\D \ln k_\mathrm{h}}{\D t} \mathcal{F}_\mathcal{G}^{(n)} - \frac{k_\mathrm{h}}{a}\left(\mathcal{F}_\mathcal{E}^{(n+1)} - \mathcal{F}_\mathcal{B}^{(n+1)}\right) - \frac{\beta}{\MP} \dot{\varphi} \mathcal{F}_\mathcal{B}^{(n)}= S_{\mathcal{G}}^{(n)}\, , 
    \end{equation}
    \begin{equation}
         \frac{\D}{\D t} \mathcal{F}_\mathcal{B}^{(n)} + (4+n)\frac{\D \ln k_\mathrm{h}}{\D t} \mathcal{F}_\mathcal{B}^{(n)} - 2\frac{k_\mathrm{h}}{a}\mathcal{F}_\mathcal{G}^{(n+1)}  =  S_{\mathcal{B}}^{(n)}\, .
    \end{equation}
    \begin{equation}
        H^2 = \frac{1}{3 \MP^2} \left(\frac{1}{2} {\dot\varphi}^2 + V(\varphi) + \frac{1}{2}  \left(\frac{a}{k_\mathrm{h}}\right)^4  ( \mathcal{F}_\mathcal{E}^{(0)} +  \mathcal{F}_\mathcal{B}^{(0)} )\right) \, ,
    \end{equation}%
    \label{eq: PAI eoms}%
\end{subequations}%
\noindent where $V(\varphi)$ is the inflaton potential, the cutoff scale is $k_\mathrm{h} =  {\max}_{{s \leq t}}(2 a(s)H(s) |\xi(s)|)$,\footnote{
    Here, $k_\mathrm{h}$ takes the role of $k_\mathrm{UV}$ in Eq.~\eqref{eq: GEF ODEs}. The renaming reflects the typical name adopted in the literature.}
 and $\xi = \beta \dot{\varphi}/ (2 H\MP)$ is the instability parameter of axion inflation.

\subsubsection*{Configuring a GEF model}

Let us see how we can use the \geff\ to solve this system. We start out by retrieving the appropriate pre-defined model using the \code{compile\_model} function.\smallskip

\begin{lstlisting}[language=mypython, numbers=left, mathescape, label=Code:getting PAI, xleftmargin=2.em]
from geff import compile_model

# Create the GEF model of our choice
paiGEF = compile_model("pai")
\end{lstlisting}
The object \code{paiGEF} is the compiled version of the \code{pai} model, and defines a solver configured to solve Eq.~\eqref{eq: PAI eoms}. To initialize \code{paiGEF}, we need to specify the setup which we want to study:
First, \code{paiGEF} expects initial data for the inflaton field, $\varphi(t=0)$ and $\dot\varphi(t=0)$. Second, we need to specify the inflaton potential, $V(\varphi)$, and the inflaton--vector coupling $\beta$.
Besides this, the model assumes that all quantities $\mathcal{F}_\mathcal{X}^{(n)}$ are negligible at $t=0$, and are thus set to zero.
To learn about the expected input, use \code{paiGEF.print\_input()}. A list of all variables, constants, and functions recognized by \code{paiGEF} can be printed by calling \code{paiGEF.print\_ingredients()}.
In both cases, the output also contains a concise description of what each quantity represents.

For our example, we configure the model to start on the slow-roll attractor of a chaotic inflation potential, about $60$ $e$-folds before the end of inflation:
$$ \varphi(0) = 15.55 \MP\,, \quad \dot{\varphi}(0) = -\sqrt{\frac{2}{3}} m \MP \,, \quad V(\varphi) = \frac{1}{2}m^2 \varphi^2\,,$$
and we set $m = 6.16 \times 10^{-6} \MP$, $\beta = 15$.

Realizing this setup using the \geff\ looks as follows:\footnote{
    All pre-defined models in the GEF use Planck units, $\MP = (8 \pi G)^{-1/2} = 1$.}\smallskip
\begin{lstlisting}[language=mypython, numbers=left, mathescape, xleftmargin=2.em]
import numpy as np

# define constants
m = 6.16e-6
beta = 15

# define initial conditions
phi = 15.55
dphi = -np.sqrt(2/3)*m

# define potential and its derivative
def V(x): return 0.5 * m**2 * x**2
def dV(x): return m**2 * x

# initialize paiGEF
mod = paiGEF(beta=beta, phi=phi, dphi=dphi, V=V, dV=dV)
\end{lstlisting}
The \texttt{mod} object is now ready to compute the inflationary dynamics of our system.

\subsubsection*{Solving the ODEs}
To solve Eq.~\eqref{eq: PAI eoms}, we use the \code{run} method of \code{mod}:\smallskip
\begin{lstlisting}[language=mypython, numbers=left, mathescape, label=code:run, xleftmargin=2.em]
sol, spec, info = mod.run()
\end{lstlisting}
The \code{run} method returns three objects: the solution to the ODEs in Eq.~\eqref{eq: PAI eoms} is contained in \code{sol},
 the evolution on the gauge-field mode functions, $A_\lambda(t,k)$, is computed and returned as the object \code{spec}, while
 \code{info} is just a byproduct that contains full information on the ODE solution. For basic applications, all information we actually want is in \code{sol} and \code{spec}.\\

The object \code{sol} stores information on all relevant variables for the background evolution. These can be accessed as attributes of \code{sol}.
For instance, the example below shows how to plot the time evolution of the energy-densities using \code{sol}:\footnote{
    We use \code{matplotlib} for illustration, but it is not required for the \geff\ and needs to be installed separately.
}
\begin{lstlisting}[language=mypython, numbers=left, mathescape, label=Code:using sol, xleftmargin=2.em]
import matplotlib.pyplot as plt

# Plot the evolution of the inflaton amplitude as a function of e-folds:
plt.plot(sol.N, sol.dphi**2/(6*sol.H**2)) # inflaton kinetic energy density
plt.plot(sol.N, (sol.E + sol.B)/(6*sol.H**2)) # electromagnetic energy density
plt.plot(sol.N, sol.V(sol.phi)/(3*sol.H**2)) # inflaton potential energy density

plt.yscale("log")
plt.ylim()

plt.show()
\end{lstlisting}
Note how \code{sol} owns several attributes \code{N}, \code{phi}, etc., which correspond to $e$-folds, inflaton amplitude, etc.
We can just use them as arrays of time (or functions in the case of \code{V}).
The \code{sol} object is a custom class used by the \code{GEFF} called \code{BGSystem}. 
It handles all relevant information on constants, variables, and functions  of a given GEF model like \code{pai}.
A \code{BGSystem} returned by an instance of the \code{pai} model (like \code{sol}) will own the following attributes:
\begin{center}
    \begin{tabular}{| l | c c c c | c c c  c| c c c | c c | c |   }
    \hline
    type & \multicolumn{4}{c|}{spacetime} & \multicolumn{4}{c|}{inflaton} & \multicolumn{3}{c|}{gauge field} & \multicolumn{2}{c|}{auxiliary} & const. \\
    \hline
    attr. & \code{t} & \code{a} & \code{N} & \code{H} & \code{phi} & \code{dphi} & \code{V} & \code{dV} & \code{E} & \code{B} & \code{G} & \code{kh} & \code{xi} & \code{beta} \\
    symbol & $t$ & $a$ & $N$ & $H$ & $\varphi$ & $\dot{\varphi}$ & $V$ & $V_{,\varphi}$ &$\langle\bm{E}^2\rangle$ & $\langle\bm{B}^2\rangle$ & $-\langle\bm{E}\cdot\bm{B}\rangle$ & $k_\mathrm{h}$ & $\xi$ & $\beta$  \\[2pt]
    \hline
    \end{tabular}
\end{center}
This information can also be found in the \href{https://richard-von-eckardstein.github.io/GEFF}{documentation} for the \code{pai} model, or accessed using \code{paiGEF.print\_ingredients()}.
The objects tracked by \code{sol} are all relevant time-dependent variables for \code{pai}, but \code{sol} also stores information on the constant $\beta$, and the inflaton potential $V$.
Note that $V$ is a function, and not a variable or constant; we can call \code{V} with \code{phi} to compute $V(\varphi)$.
To determine what any quantity \code{X} represents, one can use \code{X.get\_description()}.\\

The other important object that was returned by \code{run} is \code{spec}. This is another custom class used by the \geff\ called \code{GaugeSpec}. A \code{GaugeSpec} stores information on $A_\lambda(t,k)$ and their time derivative.
This information is used in three ways by the \geff: Firstly, we can use $A_\lambda(t,k)$ to estimate the error of our background solution by comparing $\mathcal{F}_{\mathcal{X}}^{(0)}$ as it is given in \code{sol}
to the result of integrating $A_\lambda(t,k)$ as in Eq.~\eqref{eq: GEF quantities}. 
Second, if the two evaluations of $\mathcal{F}_{\mathcal{X}}^{(0)}$  do not agree, we can use $A_\lambda(t,k)$ to recompute $\mathcal{F}_{\mathcal{X}}^{(n)}$ at some time,
and try to self-correct the ODE solution. The \code{run} method does these two things internally. Third, information on $A_\lambda(t,k)$ can be  processed further, e.g., to compute the gauge-field induced contribution
of the tensor power spectrum. \\

Both \code{BGSystem} and \code{GaugeSpec} come with their own methods to store their respective results:\smallskip
\begin{lstlisting}[language=mypython, numbers=left, mathescape, xleftmargin=2.em]
# some dummy paths for illustration
gefpath = "some_gef_file.dat"
mbmpath = "some_mbm_file.dat"

sol.save_variables(gefpath)
spec.save_spec(mbmpath)
\end{lstlisting}
The data can be restored from these files using\smallskip
\begin{lstlisting}[language=mypython, numbers=left, mathescape, xleftmargin=2.em]
from geff import GaugeSpec

sol = mod.load_GEFdata(gefpath)
spec = GaugeSpec.read_spec(mbmpath)
\end{lstlisting}
Note that \code{save\_variables} did not store information on constants or functions, so, to retrieve the full information on our GEF run, we need to use an appropriately configured instance of \code{paiGEF}, in our case \code{mod}.\footnote{
    We plan to improve on this for version 1.
}

\subsubsection*{Using the \code{geff.tools} module}
Now that we have determined the inflationary dynamics, we can use the \code{geff.tools} module.
Below, we show how to use \code{sol} and \code{spec} to first compute the tensor power spectrum for our setup, before using it to determine the resulting gravitational-wave spectrum.

\begin{lstlisting}[language=mypython, numbers=left, mathescape, xleftmargin=2.em]
from geff.tools import PowSpecT, omega_gw

# Use sol to initialize the PowSpecT class
pt_fai = PowSpecT(sol)

# Compute the vacuum and induced power spectrum for 100 momentum modes k using spec:
ks, pt_spec =  pt_fai.compute_pt(100, spec)

# from the power spectrum, we can then deduce the gravitational-wave spectrum:
f, gwspec = omega_gw(ks, pt_spec["tot"], Nend=sol.N[-1], Hend=sol.H[-1])

\end{lstlisting}
It is pretty simple; \code{sol} knows everything about the inflationary dynamics, and we extract the necessary information using the \code{PowSpecT} class.
 We can then compute the tensor power spectrum by passing \code{spec}.
 For the gravitational-wave spectrum, we need additional information about the end of inflation: the scale factor, $a_\mathrm{end} = \exp(N_\mathrm{end})$, and the Hubble rate, $H_\mathrm{end} = H(N_\mathrm{end})$.\footnote{
    The Hubble rate should be passed in Planck units, to match \code{ks}, which is also given in Planck units.
 }
 To retrieve these values, we use \code{sol.N} and \code{sol.H} like an ordinary array.\footnote{
    By default, all models in the \geff\ attempt to solve the ODEs until the end of inflation has been reached. This can be deactivated by accessing a models \code{GEFSolver} before using \code{run}. 
    For details, see the \href{https://richard-von-eckardstein.github.io/GEFF}{documentation}.}

\subsubsection*{Other models}
Besides the PAI model, also two versions of FAI are implemented in the \geff. These two versions are called \code{fai\_basic} and \code{fai\_kh} and differ in the 
precise effective treatment of the fermions. The \href{https://richard-von-eckardstein.github.io/GEFF}{documentation} gives more details, and the main physical arguments are outlined in Ref.~\cite{eckardstein_2025}.
Here, we simply illustrate how to initialize these models using the \geff, taking as an example \code{fai\_kh}.\smallskip
\begin{lstlisting}[language=mypython, numbers=left, mathescape, xleftmargin=2.em]
# initialize the model
faiGEF = compile_model("fai_kh", {"picture":"electric"})

# chose initial conditions
model = faiGEF(beta=...)

# solve the ODEs as before
sol, spec, info = model.run(...)

...
\end{lstlisting}
Note the small difference with respect to initializing \code{pai}; besides the model name, we pass a dictionary to \code{compile\_model}. This dictionary passes internal specifications of the \code{fai\_kh} model which are called ``pictures''.
These are again related to modelling assumptions on Schwinger pair production detailed in Ref.~\cite{eckardstein_2025}. Besides this, once the model has been compiled, everything else works the same.
If no settings dictionary is passed, the model will use default configurations.

Details on all pre-defined models are given in the \href{https://richard-von-eckardstein.github.io/GEFF}{documentation} including instructions on how to define a new custom GEF model.

\subsection{The machinery of the \geff}
\label{sec: inner workings}

Before concluding this tour of the \geff, we want to briefly discuss the different components of a GEF model, and how the \code{run} method interacts with them. 

As we have seen in the previous section, the \code{compile\_model} function creates an ODE solver from a GEF model, which can be executed using the \code{run} method.
The ODE solver for a given GEF model is made up of two components, the \code{GEFSolver} and the \code{ModeSolver}.
The first determines the inflationary background evolution (\code{sol} and \code{info} in the previous section), the latter computes the mode functions $A_\lambda(t,k)$ (\code{spec} in the previous section).
The two results can be compared to each other to assess the convergence of the result.

\begin{figure}[h!]
    \centering
    \includegraphics[width=0.95\textwidth]{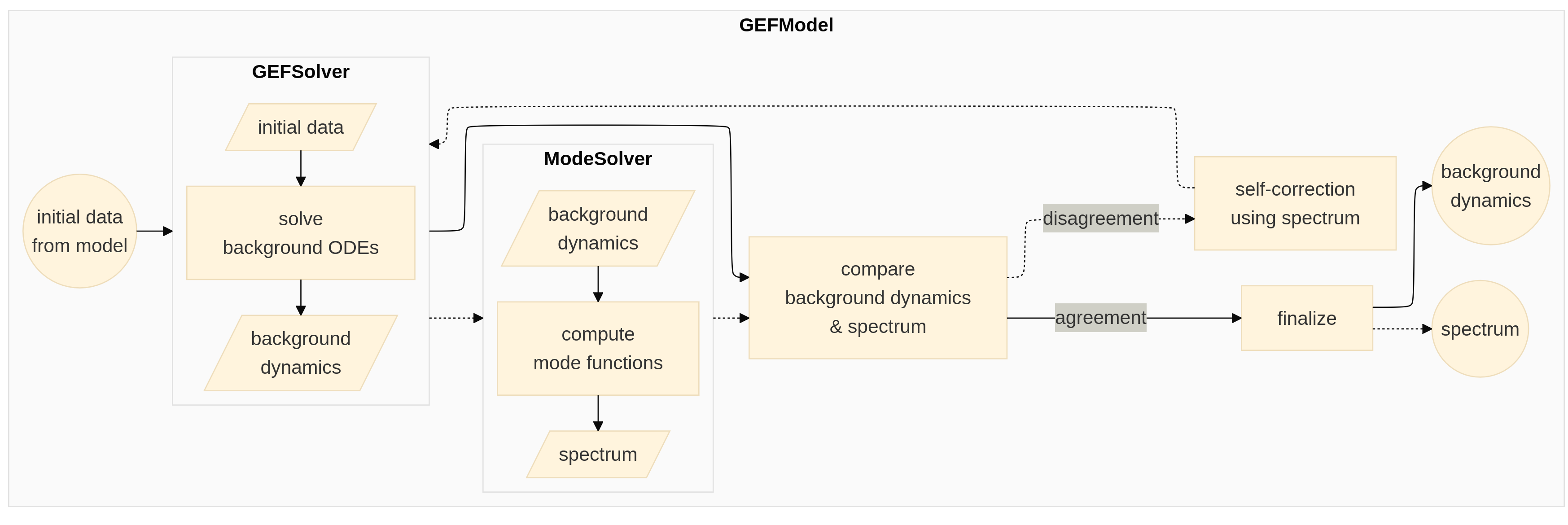}
    \caption
    { A sketch of the algorithm behind the \code{run} method}
    \label{fig: run algorithm}
\end{figure}

The algorithm used by \code{run} is sketched in Figure~\ref{fig: run algorithm}.
First, the \code{GEFSolver} computes the background evolution, which is then passed to the \code{ModeSolver} for crosschecks.
If the two results agree, the \code{run} method terminates and returns the output. Otherwise, the GEF will attempt to self-correct, using $A_\lambda(t,k)$ to compute new initial condition starting from some time $t_\mathrm{reinit}$.
The cycle is then repeated with this new input: the \code{GEFSolver} solves the ODEs, the \code{ModeSolver} computes the spectrum, and then both results are again compared to each other.\footnote{
    This is only repeated for a given number of times determined by the \code{mbm\_attempts} argument of \code{run}.
}

Note that \code{run} can be executed with the argument \code{nmodes=None}. In this case, the gauge-field spectrum is not computed, and the dotted lines in Fig.~\ref{fig: run algorithm} are passed over. This means that the background solution is not verified by the gauge spectrum.
We do not advise this outside of testing. In any case, one can always compute the spectrum later on by initializing the \code{ModeSolver} with the \code{sol} output of \code{run}.

For more details on the \code{GEFSolver} and \code{ModeSolver} see the \href{https://richard-von-eckardstein.github.io/GEFF}{documentation}, where we also discuss how to define a custom  \code{GEFSolver} and \code{ModeSolver} to realize a new GEF model.

\section{Picture gallery}
\label{sec: pictures}

\begin{figure}
    \centering
    \includegraphics[width=1\textwidth]{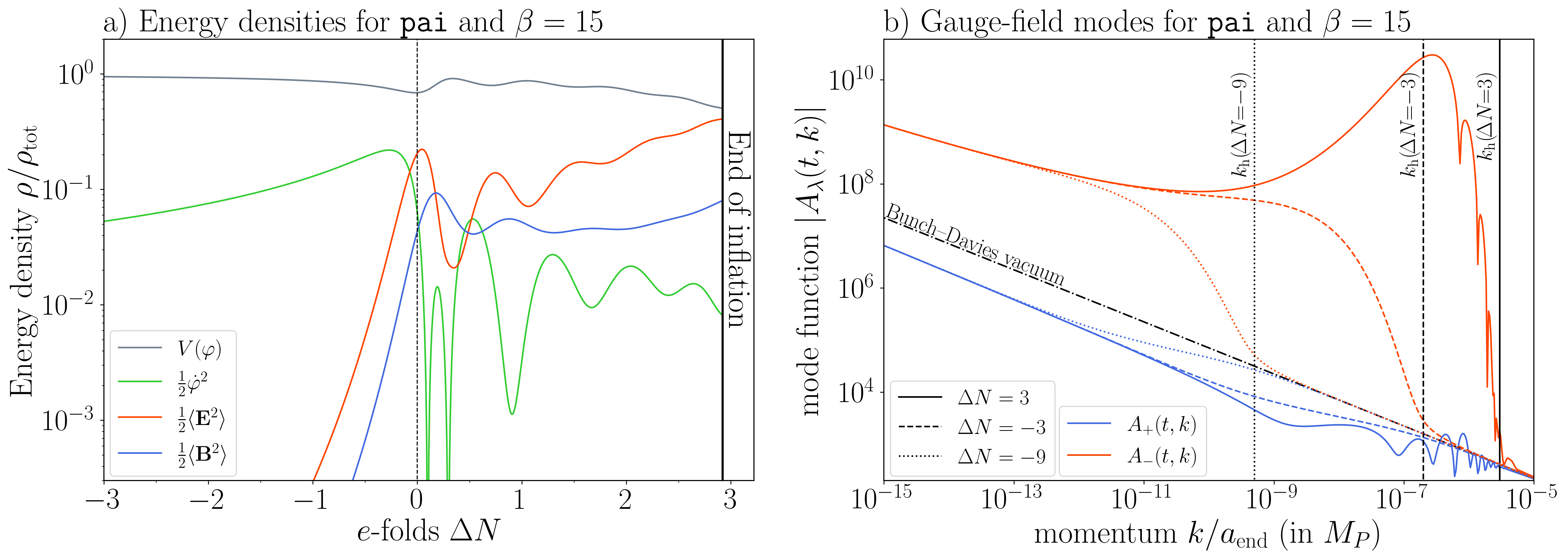}
    \caption
    { Results from \code{run} using \code{pai} with $\beta=15$ and $m=6.16\times 10^{-6} \MP$. Panel a): The evolution of the energy densities extracted from \code{sol}.
    Panel b): The gauge-mode spectra $|A_\lambda(t,k)|$ extracted from \code{spec} shown for $\Delta N = -9$, $-3$, and $3$.
    In both plots, we use $\Delta N$ as the number of $e$-folds starting from the end of inflation expected from slow-roll dynamics.}
    \label{fig: evolution}
\end{figure}

\begin{figure}
    \centering
    \includegraphics[width=1\textwidth]{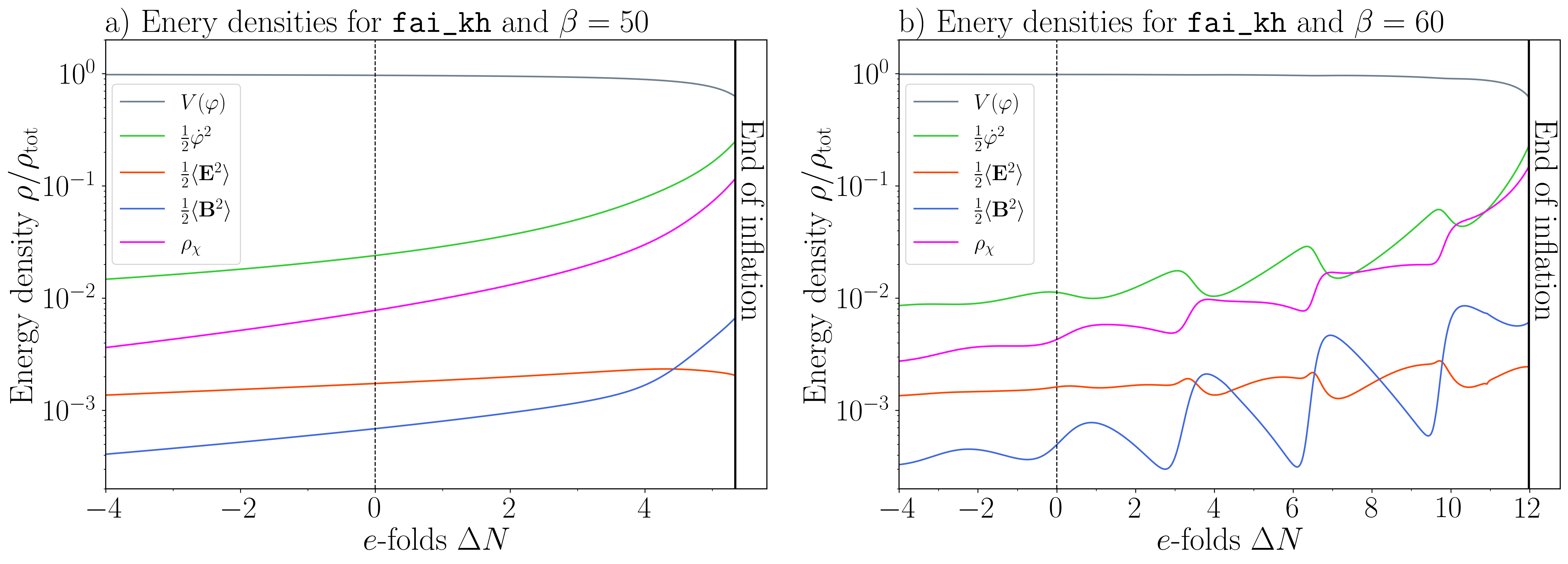}
    \caption
    {  Results from \code{run} using \code{fai\_kh} with $m=2\times 10^{-5} \MP$. Panel a): Energy-density evolution for $\beta=50$.
    Panel b): Energy-density evolution for $\beta=60$.
    $\Delta N$ is defined as in Fig.~\ref{fig: evolution}.}
    \label{fig: fai-evol}
\end{figure}

\begin{figure}
    \centering
    \includegraphics[width=1\textwidth]{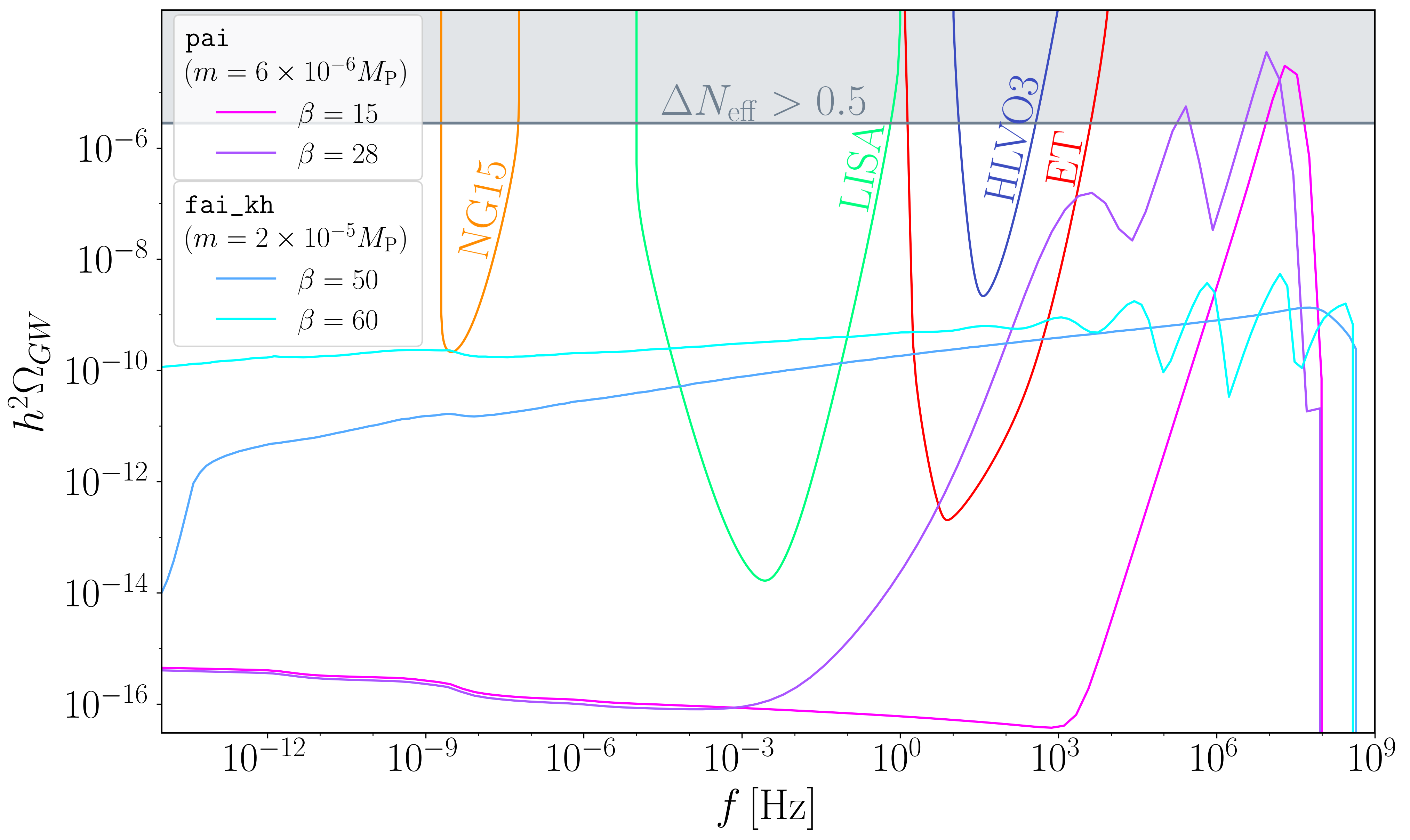}
    \caption
    { Gravitational-wave spectra for the results of Figs~\ref{fig: evolution} and~\ref{fig: fai-evol}. We also show the spectrum for \code{pai} with $\beta=28$ and $m=6.16\times 10^{-6} \MP$.}
    \label{fig: spec}
\end{figure}

To finish this introduction to the \geff\, we want to showcase some results obtained with it. 
In Fig.~\ref{fig: evolution}, we show what can be achieved with only our simple tutorial in Sec.~\ref{sec: tour}.
Some examples using \code{fai\_kh} are shown in Fig.~\ref{fig: fai-evol}. We show a variety of gravitational-wave spectra computed with the \geff\ in Fig.~\ref{fig: spec}.

\section{Conclusion}

The \geff\ --- the \textit{Gradient Expansion Formalism Factory} --- is a flexible and easy-to-use python package that is designed to study gauge-field production during inflation, a rich topic with many enticing phenomenological applications.

The package relies on the gradient expansion formalism (GEF) to capture the nonlinear dynamics of gauge-field backreaction. With this technique, one can numerically determine inflationary dynamics including gauge-field production within a few minutes,
and use this information to also compute the corresponding gauge-field mode spectrum.

Using the \geff, the user has access to pre-defined models of Abelian axion inflation. The first model is pure axion inflation (or PAI), 
which involves the axion--inflaton field directly coupled to an Abelian gauge field via the interaction $\varphi  F_{\mu\nu} \tilde{F}^{\mu\nu}$. The second model is fermionic axion inflation (FAI),
an extension of PAI assuming the Standard Model (SM) hypercharge field which sources SM fermions via Schwinger pair creation. Note that, in both models, the axion is assumed to be a homogeneous background field.
In this sense, the \geff\ serves as a precursor for lattice studies that can go beyond this assumption, but at the price of a significantly increased computational cost.
Beyond these pre-defined models, the \geff\ also allows the user to implement their own GEF models, e.g., for alternative scenarios of axion inflation, or dilatonic models.

The \geff\ also comes with tools allowing the user to take their GEF results and compute physically relevant quantities, such as the gravitational-wave spectrum induced by gauge fields.

To discover more about the \geff, please read our \href{https://richard-von-eckardstein.github.io/GEFF}{documentation}. 

\vskip.25cm
\section*{Acknowledgements}
I would like to thank my longtime collaborators Kai Schmitz and Oleksandr Sobol; it is to large parts my collaboration with them, which has lead to the development of the \geff.
Furthermore, I thank Justus Kuhlmann for advising me on the finer details of creating a \code{Python} package.

\bibliographystyle{apsrev4-1}
\bibliography{bibliography.bib}
\end{document}